\documentclass[preprint,onecolumn]{elsarticle}
\usepackage{amssymb}
\usepackage{latexsym}
\usepackage[english]{babel}
\usepackage{times,mathptmx}
\usepackage{graphicx}

\begin{document}

\begin{frontmatter}

\title{Discussion of Three Examples to Recent Results of Finite- and Fixed-Time Convergent Algorithms}

\author{Michael Basin\fnref{label2}}
\address{School of Physical and Mathematical
Sciences, Autonomous University of Nuevo Leon, San Nicolas de los
Garza, Nuevo Leon, Mexico, and ITMO University, St. Petersburg,
Russia.} \ead{mbasin@fcfm.uanl.mx}


\author{Pablo Rodriguez-Ramirez}
\address{School of Physical and Mathematical
Sciences, Autonomous University of Nuevo Leon, San Nicolas de los
Garza, Nuevo Leon, Mexico.} \ead{pablo.rodriguezrm@uanl.edu.mx}

\fntext[label2]{The authors thank the Mexican National Science and
Technology Council (CONACyT) for financial support under Grant
250611.}

\end{frontmatter}

\textbf{Abstract.} This note discusses three examples given in the
recent technical correspondence paper \cite{Seeber2019}, which
addresses the results presented in \cite{Basin2014,IET2,JFI2016}. It
is shown that the first example (\cite{Seeber2019}, Section 3) is
irrelevant to the results of \cite{Basin2014}. The second example
(\cite{Seeber2019}, Section 4) establishes a well-known fact that a
continuous differentiator can exactly differentiate a signal, only
if its second derivative is equal zero. This note provides a method
to extend the algorithms presented in \cite{IET2} to the general
case. Finally, the third example (\cite{Seeber2019}, Section 5)
presents a particular case related to Theorem 1 of \cite{JFI2016}.
Theorem 1 of \cite{JFI2016} remains, however, valid in the most
practical case of selecting control gains. The result of Theorem 2 in 
\cite{JFI2016} estimating the fixed convergence time holds as well.

\section{Introduction}

The recently published technical correspondence paper
\cite{Seeber2019} addresses the results presented in
\cite{Basin2014,IET2,JFI2016}. Three examples questioning validity
of the obtained results are provided.

This note discusses the examples given in \cite{Seeber2019}. It is
shown that the first example (\cite{Seeber2019}, Section 3) is
irrelevant to the results of \cite{Basin2014}. The second example
(\cite{Seeber2019}, Section 4) establishes the well-known fact about
the result of \cite{IET2} that a continuous differentiator can
exactly differentiate a signal, only if its second derivative is
equal zero. This note provides a method to extend the algorithms
presented in \cite{IET2} to the general case. It is shown that if
the signal second derivative is not equal to zero, the
differentiator can be modified by including discontinuous terms to
achieve the goal. Finally, the third example (\cite{Seeber2019},
Section 5) presents a special case of control gains and initial
conditions, where the result of Theorem 1 in \cite{JFI2016} on
estimating the finite convergence time does not hold. However, this
note demonstrates that the result of Theorem 1 in \cite{JFI2016}
still remains valid in the most practical case of selecting control
gains. 
The result of Theorem 2 in
\cite{JFI2016} on estimating the fixed convergence time holds as
well. 

This note is organized as follows. Section 2-4 subsequently discuss
the examples given in Sections 3, 4, and 5 of \cite{Seeber2019}.
Section 6 summarizes the discussions.

\section{Discussion of Example of Section 3 in \cite{Seeber2019}}

Following the notation of Lemma 1 in \cite{Seeber2019}, note that
the paper \cite{Basin2014} considers only systems with initial
conditions in the form $[0,x_{20},x_{30},0]$. Therefore, Lemma 1 of
\cite{Seeber2019} is applicable to the systems studied in
\cite{Basin2014}, only if $a=b=0$, that is, $x(t_0)=[0,0,0,0]$ is
the origin. However, in this case, the system (2) of Section 3 in
\cite{Seeber2019} has only the zero solution, $x(t)=0$, for $t\ge
0$, which is finite-time convergent to the origin. Remark 2 and Fig.
1 of Section 3 in \cite{Seeber2019} are not relevant to the result
of \cite{Basin2014}, since $k_3^2 =1<2$ and the conditions of Lemma
4 in \cite{Basin2014} do not hold. Proposition 3 of Section 3 in
\cite{Seeber2019} is not relevant to the result of \cite{Basin2014}
as well, since the paper \cite{Basin2014} studies only attractivity
(convergence) problems but not finite-time stability ones. The
difference between finite-time stability and finite-time
attractivity concepts can be consulted in Section 4 of \cite{JFIPF}.

\section{Discussion of Example of Section 4 in \cite{Seeber2019}}

This is a well-known fact that the finite- and fixed-time convergent
differentiators proposed in Theorems 1 and 2 of \cite{IET2} converge
to the real system states exactly, only if the output $n$-th
derivative is equal to zero, $y^{(n)}(t)=0$, for all $t\ge T$, where
$T$ is a certain finite time and $n$ is the dimension of the
differentiator. Note that in the series of papers mentioned in
\cite{Seeber2019} the finite- or fixed-time convergent
differentiators are used as parts of finite- or fixed-time
convergent controllers, whose setpoints are represented by
equilibria, that is, the condition $y^{(n)}(t)=0$ holds after a
certain finite $T$.

Furthermore, those differentiators can be modified to achieve
finite- or fixed-time convergence in the general situation by adding
the term $-\lambda sign(z_1(t)-y(t))$ to the last differentiator
equations, where $\lambda > \mid y^{(n)}(t)\mid$ is a uniform bound
for the output $n$-th derivative. For example, in case of the
fixed-time convergent differentiator proposed in Theorem 2 of
\cite{IET2}, the corresponding equations take the form
$$
\dot{z}_1(t) = z_2(t) -  k_1\mid z_1(t)-y(t)\mid
^{\alpha_1}sign(z_1(t)-y(t))
$$
$$
- \kappa_1 \mid z_1(t)-y(t)\mid ^{\beta _1}sign(z_1(t)-y(t)), \eqno
(1)
$$
$$\vdots$$
$$
\dot{z}_i(t) = z_{i+1}(t) - k_i\mid z_1(t)-y(t)\mid
^{\alpha_i}sign(z_1(t)-y(t))
$$
$$
-\kappa_i \mid z_1(t)-y(t)\mid ^{\beta _i}sign(z_1(t)-y(t)),
$$
$$
\quad i=1,\dots,n-1
$$
$$\vdots$$
$$
\dot{z}_n(t) = -k_n \mid z_1(t)-y(t)\mid ^{\alpha_n}sign(z_1(t)-y(t))
$$
$$
-\kappa_n \mid z_1(t)-y(t)\mid ^{\beta _n}sign(z_1(t)-y(t))
$$
$$
-\lambda sign(z_1(t)-y(t)),
$$
$$
k_1,\dots,k_n,\lambda > 0,
$$
where the gains $k_1,\dots,k_n$ and $\kappa_1,\dots,\kappa_n$ satisfy the conditions of Theorem 2
of \cite{IET2} and $\lambda > \mid y^{(n)}(t)\mid$. This
modification keeps the convergence fixed time estimates given in
Theorem 2 of \cite{IET2} and the convergence finite time estimates
given in Theorem 1 of \cite{IET2} in the most practical case of
selecting the control gains, as noted in the next section.

The differentiator (1) is not smooth; however, a smooth
differentiator for $(n-1)$-th derivative of the output can be
constructed by increasing the dimension of the differentiator (1) by
one, i.e., adding the equation for $z_{n+1}$ and moving the term
$-\lambda sign(z_1(t)-y(t))$ to this equation, provided that the
condition $\lambda > \mid y^{(n+1)}(t)\mid$ holds.


\section{Discussion of Example of Section 5 in \cite{Seeber2019}}

\subsection{The result of Theorem 1 in \cite{JFI2016} remains valid in the most practical case or after imposing an additional condition}

Indeed, the result of Theorem 1 in \cite{JFI2016} remains valid in
the most practical case of selecting the control gains
$k_1,k_2,\ldots,k_n$ by assigning the eigenvalues of the matrix $A$
as the multiple roots of its characteristic polynomial in the form
$(\lambda - \mu_i)^n=0$, where all $\mu_i = -\mu $ and $\mu >0 $ is
a positive real number. This is the assignment scheme mostly used by
control scientists and engineers, which is commonly implemented due
to its simplicity and the fact that increasing the absolute value of
$\mu $ leads to accelerating the convergence of a linear system
state towards the origin.

To see this, consider the example given in Section 5 of
\cite{Seeber2019}. Then, $k_1=\mu^2$, $k_2=2\mu$, and the condition
required by Remark 9 and Proposition 10 of \cite{Seeber2019} does
not hold, since $k_2^2-4k_1=4\mu^2 -4\mu^2=0$. Further calculations
yield that the corresponding matrix $P$ is given by
$$\label{P}
P =\left(
         \begin{array}{cc}
           \frac{1}{\mu} + \frac{\mu^2 +1}{4\mu^2} & \frac{1}{2\mu^2}\\
           \frac{1}{2\mu^2} & \frac{\mu^2 +1}{4\mu^3}
         \end{array}
       \right).
$$
The right-hand side of the formula (22) in \cite{Seeber2019} takes
the form $ \mu \frac{\lambda_{max}(P)}{\lambda_{min}(Q)}$. Assuming
$\lambda_{min}(Q)=1$, the inequality $\mu \lambda_{max}(P)>1$ holds
for any $\mu>0$, which is verified directly calculating the maximum
eigenvalue of the matrix $P$ as a function of $\mu $. For instance,
$\mu \lambda_{max}(P)=1+\frac{\sqrt{2}}{2}$, if $\mu =1$.

Thus, Theorem 1 in \cite{JFI2016} still provides a method to
estimate the finite convergence time for Bhat and Bernstein
algorithm \cite{BB05} in the most practical and broadly employed
case of selecting its control gains $k_1,k_2,\ldots,k_n$. Validity of the formula (6) of Theorem 1 in
\cite{JFI2016} in this case is illustrated by the following
simulations.

%
%
%

The $n$-dimensional chain of integrators
$$
\dot{x}_1(t) = x_2(t), \quad x_1(t_0)=x_{10},
$$
$$
\dot{x}_2(t) = x_3(t), \quad x_2(t_0)=x_{20},
$$
$$
\cdots
$$
$$
\dot{x}_n(t) = u(t), \quad x_n(t_0)=x_{n0},
$$
is simulated for $n=2,3,4,5$. The scalar control input $u(t)$ is
assigned according to Bhat and Bernstein algorithm \cite{BB05}
$$
u(t) = v_1(t) + v_2(t) + \ldots + v_n(t),
$$
where $v_i(t)=-k_i \mid x_i(t)\mid^{\gamma_i}sign(x_i(t))$ and the
exponents $\gamma _i$, $i=1,\ldots,n$, are defined by $\gamma
_{i-1}=\gamma _i\gamma _{i+1}/(2\gamma _{i+1}-\gamma _i)$,
$i=2,\ldots,n$, $\gamma _{n+1}=1$, and $\gamma _{n}=\gamma$. The
control gains $k_i$, $i=1,\ldots,n$, are assigned such that all
multiple roots of its characteristic polynomial $(\lambda -
\mu_i)^n=0$ are equal to $\mu_i = -1$. Namely, $k_1=1$, $k_2=2$ for
$n=2$; $k_1=1$, $k_2=3$, $k_3=3$ for $n=3$; $k_1=1$, $k_2=4$,
$k_3=6$, $k_4=4$ for $n=4$; and $k_1=1$, $k_2=5$, $k_3=10$,
$k_4=10$, $k_5=5$ for $n=5$. The parameter $\gamma $ is set to
$\gamma =10/11$ in all simulations. The convergence time estimated
is computed according to the formula (6) of Theorem 1 in
\cite{JFI2016}.

The simulation results are given in the following tables, which
confirm validity of the formula (6) of Theorem 1 in \cite{JFI2016}.
\begin{center}
    \begin{tabular}{||c c c c c c||}
        \hline
        \textbf{Convergence time} & \multicolumn{5}{|c|}{\textbf{n=2}}\\
        \hline
        \hline
        {Initial Conditions $x_i(0)$} & 0.01 & 1 & 100 & 10,000 & 1'000,000 \\ [0.5ex]
        \hline
        Simulation (s)& 7.7 & 13.9 & 23.1 & 37 & 58.05.5 \\
        \hline
        Estimated Time (s)& 7.95 & 17.07 & 37.08 & 82.56 & 188.59 \\
        \hline
        Rate & 1.03 & 1.22 & 1.60 & 2.23 & 3.24 \\
[1ex]
        \hline
    \end{tabular}
\end{center}
\begin{center}
    \begin{tabular}{||c c c c c||}
        \hline
        \textbf{Convergence time} & \multicolumn{4}{|c|}{\textbf{n=3}} \\
        \hline
        \hline
        {Initial Conditions $x_i(0)$} & 1 & 100 & 10,000 & 1'000,000 \\ [0.5ex]
        \hline
        Simulation (s) & 25.33 & 39.1 & 59.7 & 90.6 \\
        \hline
        Estimated Time (s) & 46.48 & 92.99 & 209.29 & 500.43 \\
        \hline
        Rate & 1.83 & 2.37 & 3.50 & 5.52 \\
[1ex]
        \hline
    \end{tabular}
\end{center}
\begin{center}
    \begin{tabular}{||c c c c c||}
        \hline
        \textbf{Convergence time} & \multicolumn{4}{|c|}{\textbf{n=4}} \\
        \hline
        \hline
        {Initial Conditions $x_i(0)$} & 1 & 100 & 10,000 & 1'000,000 \\ [0.5ex]
        \hline
        Simulation (s) & 40.6 & 60 & 89 & 132.3 \\
        \hline
        Estimated Time (s) & 146.09 & 289.71 & 650.98 & 1528.4 \\
        \hline
        Rate & 3.59 & 4.82 & 7.31 & 11.55 \\
[1ex]
        \hline
    \end{tabular}
\end{center}
\begin{center}
    \begin{tabular}{||c c c c c||}
        \hline
        \textbf{Convergence time} & \multicolumn{4}{|c|}{\textbf{n=5}} \\
        \hline
        \hline
        {Initial Conditions $x_i(0)$} & 1 & 100 & 10,000 & 1'000,000 \\ [0.5ex]
        \hline
        Simulation (s) & 60.8 & 87.4 & 126.6 & 185.1 \\
        \hline
        Estimated Time (s) & 508.21 & 998.29 & 2208.8 & 5108.2 \\
        \hline
        Rate & 8.35 & 11.42 & 17.44 & 27.59 \\
[1ex]
        \hline
    \end{tabular}
\end{center}

The authors thank the author of \cite{Seeber2019} for the example
given in Section 5 of \cite{Seeber2019} as the really relevant and
insightful one.

\subsection{The result of Theorem 2 in \cite{JFI2016} remains valid}

It is argued in Subsection 5.3 of \cite{Seeber2019} that the
inequality (23) there is not valid for all $x\in R^n$, since both
parts of the inequality (23) tend to zero as $x$ tends to zero.
Following this logic, the example of Subsection 5.2 could be
constructed only for initial values $x_0$ sufficiently close to
zero. Furthermore, the result of Theorem 2 in \cite{JFI2016}
providing an upper estimate for fixed convergence time would remain
valid, since it takes into account initial values arbitrarily
distant from zero.

Indeed, consider the example given in Section 5 of
\cite{Seeber2019}. Let the gains $\kappa_1,\kappa_2$ in Theorem 2 in
\cite{JFI2016} are selected the same as $k_1,k_2$: $k_1=\kappa_1=1$,
$k_2=\kappa_2=6$. Then, assuming $P_1=P$ and setting $Q_1=Q$ to the
$2\times 2$ identity matrix, the right-hand side of the formula (22)
in the fixed-time convergence case is equal to
$$
2\frac{k_2-\sqrt{k_2^2-4k_1}}{2}\frac{\lambda_{max}(P)}{\lambda_{min}(Q)}
=(3-\sqrt{8})(\frac{10}{3}+\sqrt{10})\approx 1.14447>1.
$$

Thus, the result of Theorem 2 in \cite{JFI2016} remains valid and,
in addition, provides a practically useful upper estimate for fixed
convergence time in the example given in Section 5 of
\cite{Seeber2019}. It should be noted that convergence time
estimates based on Lyapunov functions proposed in \cite{MF2012} are
too conservative and cannot be used for practical estimation of
fixed convergence time.

\section{Conclusions}

This note discussed the examples given in \cite{Seeber2019}. It has
been shown that the results opposed in \cite{Seeber2019} remain
valid in most practical cases or can be successfully modified or are
irrelevant to the given examples.

\section{Ackowledgments}

The authors thank Prof. Y. Shtessel, with the University of Alabama in Huntsville,
for his valuable and very useful discussions, comments, and suggestions.

\bibliography{2018ANR}{}
\bibliographystyle{elsarticle-num}

\end{document}